\documentclass[10pt,a4paper]{article} 
\usepackage{amsmath}
\usepackage[left=2.54cm,top=2.54cm,right=2.54cm,bottom=2.54cm,nohead]{geometry}
\usepackage{graphicx}
\usepackage{float}
\usepackage{setspace}
\usepackage{placeins}
\usepackage{rotating}
\usepackage{authblk}
\usepackage{xcolor}
\usepackage{nicefrac}
\usepackage[makeroom]{cancel}
\usepackage{bm}
\usepackage{textcomp}
\usepackage[colorlinks = true,
linkcolor = blue,
urlcolor  = blue,
citecolor = blue,
anchorcolor = blue]{hyperref}
\title{AC-Driven Electro-Osmotic Flow in Charged Nanopores}

\renewcommand{\thefootnote}{\fnsymbol{footnote} $,$}

\author[1]{J. Catalano\thanks{Corresponding author email: \href{mailto:jcatalano@eng.au.dk}{jcatalano@eng.au.dk}}}
\author[2]{P.M. Biesheuvel}
\affil[1]{Department of Engineering, Aarhus University, Hang\o vej 2, 8200 Aarhus, Denmark.}
\affil[2]{Wetsus, European Centre of Excellence for Sustainable Water Technology, Oostergoweg 9, 8911 MA Leeuwarden, The Netherlands.}
\providecommand{\keywords}[1]{\textbf{\textit{Keywords---}} #1}
\date{} 
\begin{document}
\maketitle

\begin{abstract}
	
	In this paper we report the theory describing the electro-osmotic flow in charged nanopores with constant radius and charge density driven by alternating current. We solve the ion and solution transport in unsteady conditions as described by the Navier-Stokes and Nernst-Planck equations considering the electrical potential inside the charged nanopore uniform in the radial direction (Uniform Potential model approximation). We derive the transport equation system in the case in which the pore is connected to two boundary diffusion layers and the cations and anions have different diffusion coefficients. This approach allows the theoretical description of the characteristic frequency dependence of the phase shift between the applied current density and the electro-osmotic flow. Additionally we show how the analysis of the dynamic response of the electro-osmotic coupling factor versus the AC frequency allows us to quantify the apparent ion diffusion coefficient in membranes. Notably, the frequency window where the phase-shift is predicted is well inside the commonly used sampling rate for electro-osmotic flow experiments $f\in\left[10^{-4},10^1\right]$ Hz, hence the proposed method can be useful to determine the apparent diffusion coefficient of ions (such as redox complex for flow batteries) in charged membranes. 
	
\end{abstract}

\keywords{Unsteady  Ion and Solution Transport; Uniform Potential Model; AC-driven Electro-Osmotic Flow; Electrokinetic Dynamic Response}
\renewcommand{\thefootnote}{\fnsymbol{footnote}}
\section{Introduction}

Transport phenomena of solutions and ions in nanoscopic charged pores and membranes are crucial in many processes in biochemistry and chemistry. \cite{COOPER198551,Hediger2004} The development of novel nano-engineered techniques sparked the interest on developing advanced materials with tailored mass transport characteristics (e.g. carbon nanotubes and grafted polymer electrolyte nanopores). \cite{Holt1034,sun2000,Regan2004CarbonNA} A subset of these nano-structured materials comprises porous structures of which the surfaces are either ionizable in the solutions used or functionalised with immobile charges (e.g. grafted polymers, polyelectrolyte layers or ion channels in polymeric matrices). In this latter case the ion transport and the coupled volumetric flow of solution due to pressure and voltage gradients is defined as electro-osmotic flow (EOF). EOF can be exploited in a variety of systems such as pH-gated or voltage-gated pores, DNA systems, drug delivery, microfluidics, and in ionic circuits.\cite{Buchsbaum2014,Whitesides2006TheOA,Xia2008}\par
EOF (in aqueous solutions) relies on the coupled interactions between the mobile ions in solution, the polar water molecules, and the (immobile or ionizable) charges at the pore walls. An electrokinetic (EK) pump makes use the EOF principle and it provides a volumetric flow of solution against a pressure gradient without any moving parts converting the applied current density into kinetic energy. The main idea is that the two surfaces of the separator are in contact with reservoirs having the same electrolyte strength (same electrolyte and concentration). An EK device is operated as a pump when the voltage-driven flow is opposite in sign and higher in magnitude with respect to the pressure-driven one.\cite{CATALANO2015}. Membrane-based EK pumps can theoretically operate at hundreds of bar of hydrostatic pressure difference and the reliable flow control at very low flow rates makes this technology viable for the construction of ion circuits and microfluidic applications even working with very dilute solutions ($< 10^{-4}$ M). The magnitude of OEF in an ionic channel network is a function of the morphological characteristics of the membrane (or pores) and markedly depends upon the characteristic pore dimension, charge density and electrolyte solution used.\cite{Haldrup2015,Haldrup2016} Thus a pore model is needed to describe the influence of the morphological parameters on the EOF efficiency. To this end the capillary or Space-Charge (SC) and the Uniform Potential (UP) models were successfully used to describe the flow characteristics in nanoscopic pores.\cite{BASU1997} The SC model is centred on the combination of the Navier-Stokes, Poisson and Nernst-Planck equations and can be derived to account for immobile or ionizible charges and fluid slip on the pore walls .\cite{MorrisonOsterle1965,GrossOsterle1968,FairOsterle1971,Sasidhar1981,Peters2016,CatalanoArxiv2016,KRISTENSEN2017,Ryzhkov2017,RYZHKOV2018} The UP model drops the radial dependence of the electrical potential in the pore as described by the Poisson equation and consider the radial potential uniform.\cite{Hawkins1989, Bowen2002, Verbrugge1990, Tedesco2016,kedem1963_I} This simplification limits the validity of the UP model to dilute solutions and narrow pore diameters; This condition is fulfilled when the electrical double layers (EDLs) inside the pores overlap. Nonetheless these conditions for highly charged membranes are of interest since the range in which EDLs overlap, extends to relatively high concentrations of $\sim$ 100 mM.\par 
Generally, an EK pump is operated with direct current (DC) and in a configuration similar to e.g. flow cells to minimize the parasitic resistances and the penalizing effects of the concentration polarization phenomena at the solution-membrane interfaces.\cite{CATALANO2014,OSTEDGAARDMUNCK2017} Instead, here we theoretically describe the operation of an EK pump operating with alternating current (AC). This latter case is not representative of any real-world EK pump since for homogeneous membranes (constant membrane morphology and charge density throughout the membrane thickness) this operation results in a nil volumetric flow (for unbiased applied voltages differences). Nonetheless the main idea is that the application of an AC signal can be a powerful tool for evaluating the membrane properties for the mass transport, in the same vein as Electrochemical Impedance Spectroscopy (EIS) is successfully used for characterizing the electrical resistances in a membrane electrode assembly.\cite{MoyaHorno1999,Moya2012}\par
Here we derive the equations for the UP model in unsteady conditions considering the nanoscopic pore in equilibrium with two boundary diffusion layers and for unequal diffusion coefficients for cations and anions. We report the characteristic phase-lag and magnitude of the EO coupling factor at different frequencies for a (highly) charged pore in dilute solution (50 mM). We demonstrate how the dynamic response of the AC-driven EOF can be successfully used to determine the coion and counterion ion diffusion coefficient in highly charged membranes.

\section{Unsteady  Electro-Osmotic Flow}
\begin{figure}
	\centering
	\includegraphics[width=0.70\textwidth, page=1]{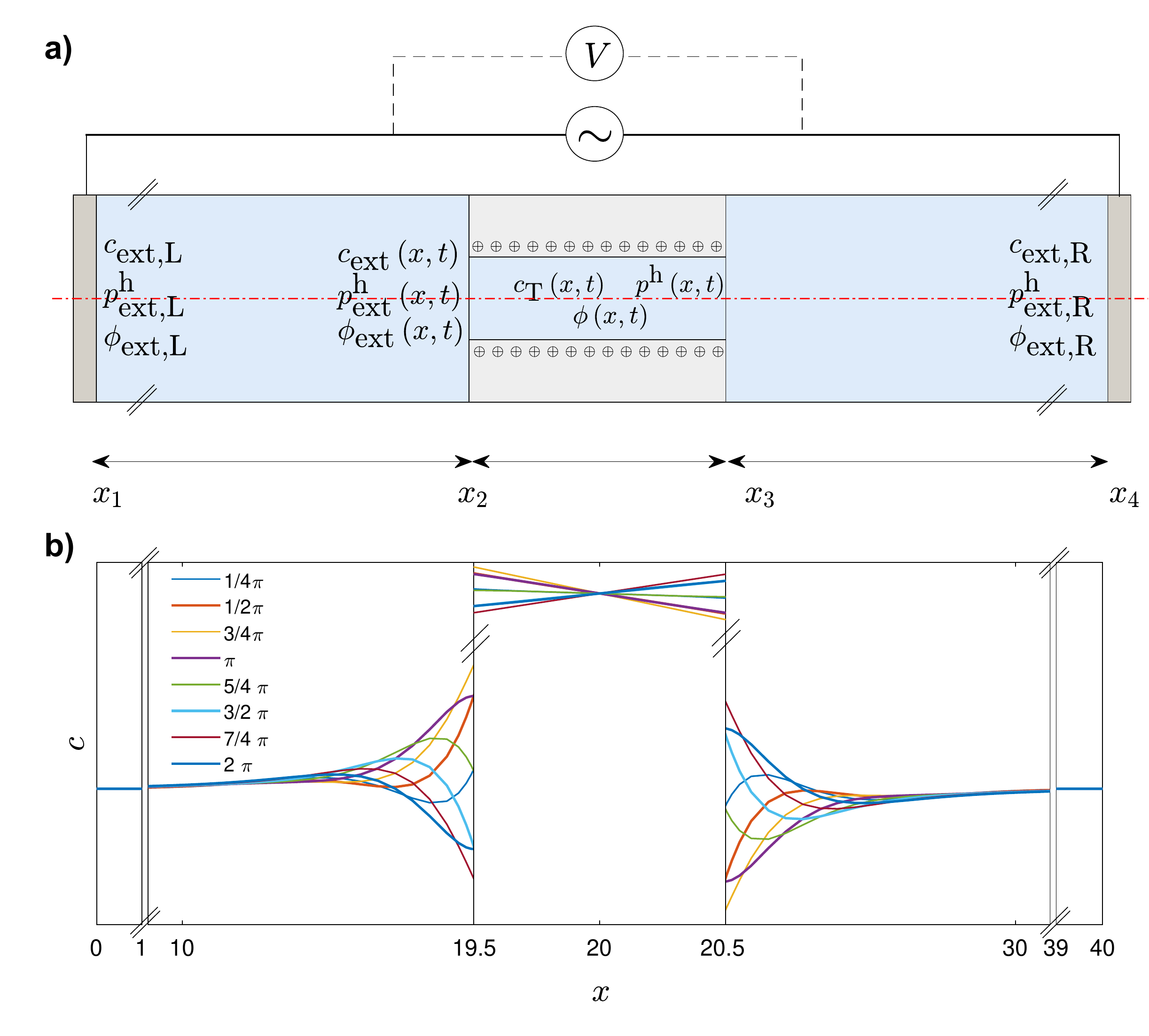}
	\caption{a): Pore geometry considered in the present work. The domains $x \in \left[x_1,x_2\right]$ and $x \in \left[x_3,x_4\right]$ are the boundary diffusion layers while the charged pore is in between the two solution-pore boundaries at $x=x_2$ and $x=x_3$. b): Example of the concentration profiles at different time frames when AC (frequency $f=1$) is applied to the electrodes.}
	\label{Fig:Schematic}
\end{figure}
With reference to Figure \ref{Fig:Schematic} let us consider a charged pore (of nanoscopic dimensions) which ends, at boundaries $x_2$ and $x_3$, are in equilibrium with two boundary diffusion layers (BDLs) having different hydrostatic pressure ($p^\textrm{h}_\textrm{ext}$), electrical potential ($\phi_\textrm{ext}$) and electrolyte concentration ($c_\textrm{ext})$. To describe a true membrane we consider an array of parallel cylindrical channels having the same surface charge density and radius. We consider the cylindrical channels evenly spaced on the membrane surface and we define the membrane porosity, $\epsilon_\textrm{p}$, as the ratio between the open area of the pores and the geometrical dimension of the membrane surface. In these conditions the system can be modelled as a nano-capillary (cross sectional area $A_\textrm{p}$) in equilibrium with the BDLs (having cross sectional area $A_\textrm{BDL}=A_\textrm{p}/\epsilon_\textrm{p}$) where we neglect the fluid-wall friction.\cite{Andersen2012}\par
To describe the unsteady-state ion and solution transport within the pore, domain $x \in \left[x_2,x_3\right]$, we start from the phenomenological equations described in Ref. \cite{MorrisonOsterle1965,GrossOsterle1968,FairOsterle1971,Sasidhar1981,Peters2016}. The force-flux equations, for an isothermal process, and considering steady-state conditions, in their dimensionless form read as:
\begin{equation} \label{eq:Phenomenological_gen}
\begin{pmatrix} u_x \\ j_\textrm{ions} \\j_{\textrm{ch}} \end{pmatrix} =
\begin{pmatrix} L_{11} & L_{12} & L_{13}\\ L_{21} & L_{22} & L_{23}\\ L_{31} & L_{32} & L_{33} \end{pmatrix}
\begin{pmatrix} -\partial p^{\textrm{t}}_\textrm{v}/\partial x \\ -\partial \mu_{\textrm{v}}/\partial x \\-\partial \phi_\textrm{v}/\partial x \end{pmatrix} 
\end{equation}
\noindent where the (radially averaged) fluxes in the \textit{x}-direction $u_x$, $j_\textrm{ions}$, and $j_\textrm{ch}$ are the volumetric flux, the total ion flux and current density respectively. The driving forces are virtual (see details in Ref. \cite{Peters2016}) and each slice of the pore is considered in equilibrium with a virtual reservoir having total pressure $p^{\textrm{t}}_\textrm{v}=p^\textrm{h}_\textrm{v}-2c_\textrm{v}$, chemical potential $\mu_\textrm{v}=\ln c_\textrm{v}$ and electrical potential $\phi_\textrm{v}$. In this work we consider the electrical potential inside the pore, $\psi$ uniform in the radial direction. This assumption is valid for narrow pores and when the electrical double layers (EDLs) inside the pore overlap, hence the Debye screening length is larger than the pore radius. This condition for highly charged membranes, such as the ones used for fuel cells, flow batteries or (reverse) electrodialysis, generally applies for external concentrations less than about 100 mM (see e.g. Supporting Information in Ref. \cite{OSTEDGAARDMUNCK2017} and \cite{Kristensen2016}). With the latter assumption the following relations hold:
\begin{equation}
\begin{aligned}
\omega X_\textrm{m}&=c_{-}-c_{+}=2c_\textrm{v} \sinh\psi\\
c_\textrm{T}&=c_{-}+c_{+}=2c_\textrm{v} \cosh\psi\\
\end{aligned}
\end{equation}   
in which $X_\textrm{m}$, $\omega$, and $c_\textrm{T}=\sqrt{X_\textrm{m}^2+\left(2c_\textrm{v}\right)^2}$ are the membrane charge density (referred to the volume of the solution inside the pore), the sign of the immobile charges ($\omega=+1$ for anion exchange membranes), and the total ion concentration, respectively. The transport coefficients $L_{ij}$, considering unequal diffusion coefficients for the anions ($D_{-}$) and cations ($D_{+}$) can be written as:
\begin{equation}\label{eq:Lmatrix_UP}
\begin{aligned}
L_{11}&=+\;\frac{1}{8\alpha} \\
L_{12}&= L_{21} = +\;\frac{1}{8\alpha}\cdot c_\textrm{T}  \\
L_{13}&= L_{31} = - \; \frac{1}{8\alpha} \omega X_\text{m} \\
L_{22}&= +\;\frac{1}{8\alpha}\cdot c_\textrm{T}^2 + 2 c_\text{v} \cosh \left(\psi^*\right) \\
L_{23}&= L_{32} = - \;\frac{1}{8\alpha} \omega  X_\text{m} \cdot c_\textrm{T} - 2 c_\text{v} \sinh \left(\psi^*\right) \\
L_{33}&=\frac{1}{8\alpha} \cdot X_\text{m}^2+2 c_\text{v} \cosh \left(\psi^*\right) \\
\end{aligned}
\end{equation}
\noindent where we note that $L_{ij}=L_{ji}$ since the system is proven Onsager symmetric. The modified radial potential $\psi^\star$, which takes into account the unequal diffusion coefficients, is defined as \cite{GrossOsterle1968,Sasidhar1981,CatalanoBiesheuvel2016}:
\begin{equation}
\psi^\star=\psi+\xi
\end{equation}
where $\xi= \nicefrac{1}{2} \ln \left( D_{-}/D_{+}\right)$. Eq. (\ref{eq:Phenomenological_gen}) can be considerably simplified when substituting the real pressures ($p^\textrm{h}$), concentrations ($c_\textrm{T}$) and electrical potentials ($\phi$) for the corresponding virtual quantities (as described in Ref. \cite{Peters2016}). Within this framework, the transport equations, now written for unsteady-state conditions, can be expressed as:
\begin{equation}
\begin{aligned}\label{Eq:Unsteady_different_D}
0&=-\frac{1}{8\alpha}\left(\frac{\partial p^\textrm{h}\left(x,t\right)}{\partial x}-\omega X_\textrm{m}\left(x\right)\frac{\partial \phi\left(x,t\right)}{\partial x}\right)-u_x\left(t\right)\\
\frac{\partial c_\textrm{T}\left(x,t\right)}{\partial t}&=-\frac{\partial j_\textrm{ions}\left(x,t\right)}{\partial x}\\
&=-u_x\left(t\right)\frac{\partial c_\textrm{T}\left(x,t\right)}{\partial x} +\cosh\xi\frac{\partial^2 c_\textrm{T}\left(x,t\right)}{\partial x^2}+\omega\sinh\xi\frac{\partial^2 X_\textrm{m}\left(x\right)}{\partial x^2}\\
&-\frac{\partial}{\partial x}\left(\left(\omega X_\textrm{m}\left(x\right)\cosh\xi+c_\textrm{T}\left(x,t\right)\sinh\xi\right)\frac{\partial \phi\left(x,t\right)}{\partial x}\right)\\
0&=-\omega X_\textrm{m}\left(x\right)u_x\left(t\right)+\sinh\xi\frac{\partial c_\textrm{T}\left(x,t\right)}{\partial x}+\omega\cosh\xi \frac{\partial X_\textrm{m}\left(x\right)}{\partial x}\\
&-\left(c_\textrm{T}\left(x,t\right)\cosh\xi+\omega X_\textrm{m}\left(x\right)\sinh\xi\right)\frac{\partial \phi\left(x,t\right)}{\partial x}-j_\textrm{ch}\left(t\right)\\
\end{aligned}
\end{equation}
and, in the case of constant charge density along the pore length, $\partial X_\textrm{m}\left(x\right)/\partial x=0$, and equal diffusion coefficient $D_\textrm{+}=D_\textrm{-}$, simplify to: 
\begin{equation}\label{Eq:UP_unsteady_d}
\begin{aligned}
0=& -\frac{1}{8\alpha}\left(\frac{\partial p^\textrm{h}\left(x,t\right)}{\partial x}-\omega X_\textrm{m}\frac{\partial \phi\left(x,t\right)}{\partial x}\right)-u_x\left(t\right)\\
\frac{\partial c_\textrm{T}\left(x,t\right)}{\partial t}=&-\frac{\partial j_\textrm{ions}\left(x,t\right)}{\partial x}\\
=& -u_x\left(t\right)\frac{\partial c_\textrm{T}\left(x,t\right)}{\partial x} +\frac{\partial}{\partial x}\left(\frac{\partial c_\textrm{T}\left(x,t\right)}{\partial x}-\omega X_\textrm{m}\frac{\partial \phi\left(x,t\right)}{\partial x}\right)\\
0=& -\omega X_\textrm{m} u_x\left(t\right)-c_\textrm{T}\left(x,t\right)\frac{\partial \phi\left(x,t\right)}{\partial x}-j_\textrm{ch}\left(t\right)\\
\end{aligned}
\end{equation}
For constant total ion flux, $\partial j_\textrm{ions} / \partial x=0$, Eq. (\ref{Eq:UP_unsteady_d}) leads to the set of equations (Eq. (53)) reported in Ref. \cite{Peters2016}. By using the UP model the only membrane morphological property besides the membrane charge density is the pore radius ($r_\textrm{p}$) which enters in the dimensionless solution viscosity ($\alpha$, dimensional form $\mu$ in Pa$\cdot$ s). The dimensionless quantities are defined as: 
\begin{equation}\label{Eq:Dimensionless_form}
\begin{aligned}
x&=\frac{x_\textrm{dim}}{L_{\textrm{ref}}}  &t&=\frac{t_\textrm{dim}}{t_\textrm{ref}} &t_\textrm{ref}&=\frac{L_\textrm{ref}^2}{D_\textrm{ref}} &c&=\frac{c_\textrm{dim}}{c_\textrm{ref}}; &\alpha=\frac{\mu D_\textrm{ref}}{p_\textrm{ref} r_\textrm{p}^2}\\
\phi&=\frac{\phi_\textrm{dim}}{\Phi_\textrm{B}}  &u_x&=\frac{u_{x,\textrm{dim}}}{u_{\textrm{ref}}} 
&\Phi_\textrm{B}&= \frac{R_\textrm{g}T}{F} &u_\textrm{ref}&=\frac{D_\textrm{ref}}{L_\textrm{ref}}&\\
p&=\frac{p_\textrm{dim}}{p_\textrm{ref}} &j&=\frac{j_\textrm{dim}}{j_\textrm{ref}} &p_\textrm{ref}&=R_\textrm{g}Tc_\textrm{ref} & j_\textrm{ref}&=\frac{D_\textrm{ref}c_\textrm{ref}}{L_\textrm{ref}}&\\
\end{aligned}
\end{equation}
\noindent where  $\Phi_\textrm{B}$, $R_\textrm{g}$, $T$, $F$, and $D_\textrm{ref}$ are the thermal voltage (in V), the gas constant (8.3144 J mol$^{-1}$ K$^{-1}$), temperature (in K), Faraday constant (96485 C mol$^{-1}$), and reference diffusion coefficient ($D_\textrm{ref}=\sqrt{D_{+}\cdot D_{-}}$ in m$^2/$s)  \cite{GrossOsterle1968,Sasidhar1981,CatalanoBiesheuvel2016}, respectively and we set the reference length and concentration to $L_\textrm{ref}=100$ $\mu$m and $c_\textrm{ref}=1$ mM, respectively. The system of equations (Eqs. (\ref{Eq:Unsteady_different_D}) and (\ref{Eq:UP_unsteady_d})) needs the proper step changes at the membrane-solution interfaces which are embedded in the following boundary conditions (at $x=x_2$ and $x=x_3\; \forall t$):
\begin{equation}\label{Eq:Up_model_BC}
\begin{aligned}
p^\textrm{h}=& p^\textrm{h}_\textrm{ext}+c_\textrm{T}-2c_\textrm{ext}\\
c_\textrm{T}=& \sqrt{X_\textrm{m}^2+\left(2c_\textrm{ext}\right)^2}\\
\phi=& \phi_\textrm{ext}+\sinh^{-1}\left(\omega X_\textrm{m}/2c_\textrm{ext}\right)\\
\end{aligned}
\end{equation}
\noindent where subscript \,''ext\,'' refers to the values of the properties in the external phase. In the following we consider two different cases: \textbf{I} constant concentration at the pore mouths, and \textbf{II} the presence of BDLs in contact with the pore.\par
\textbf{Case I:} The two pore ends are kept at constant concentration and the system can be modelled not accounting for the BDL domains. In this case the boundary conditions at $x_2$ and $x_3$ simply read as:
\begin{equation}
c_\textrm{ext}\left(x=x_2,t\right)=c_\textrm{ext,L}; \quad c_\textrm{ext}\left(x=x_3,t\right)=c_\textrm{ext,R}\\ 
\end{equation}
\noindent Here we consider the (hydrostatic) pressures of the external solution constant in time:
\begin{equation}
p_\textrm{ext}^\textrm{h}\left(x=x_2,t\right)=p_\textrm{ext,L}^\textrm{h}; \quad p_\textrm{ext}^\textrm{h}\left(x=x_3,t\right)=p_\textrm{ext,R}^\textrm{h}\\
\end{equation}
\noindent and the voltage is imposed at the solution-side of the membrane-solution interfaces ($x=x_3$ grounded $\phi(x_3)=0$).\par
\textbf{Case II:} The concentrations at the two pore ends are allowed to vary with time. In the BDLs (for $x \in \left[x_1, x_2\right]$ and $x \in \left[x_3, x_4\right]$) the following two equations are solved:
\begin{equation}\label{Eq:UP_unsteady_dvar2}
\begin{aligned}
\frac{\partial c_\textrm{ext}\left(x,t\right)}{\partial t}=&-\frac{1}{2}\frac{\partial j_\textrm{ions,ext}}{\partial x}\\
=& - u_{x,\textrm{ext}}\left(t\right)\frac{\partial c_\textrm{ext}\left(x,t\right)}{\partial x} +\frac{\partial}{\partial x}\left(\frac{\partial c_\textrm{ext}\left(x,t\right)}{\partial x}\right)\\
0=& -c_\textrm{ext}\left(x,t\right)\frac{\partial \phi_\textrm{ext}\left(x,t\right)}{\partial x}-j_\textrm{ch,ext}\left(t\right)\\
\end{aligned}
\end{equation}
while the hydrostatic pressure is invariant with \textit{x}. As boundary conditions we consider the concentrations and hydrostatic pressures at the boundaries of the two BDLs ($x=x_1$ and $x=x_4$) constant:
\begin{equation}
\begin{aligned}
c_\textrm{ext}\left(x=x_1,t\right)&=c_\textrm{ext,L};& \quad c_\textrm{ext}\left(x=x_4,t\right)&=c_\textrm{ext,R}\\ 
p_\textrm{ext}^\textrm{h}\left(x=x_1,t\right)&=p_\textrm{ext,L}^\textrm{h};& \quad p_\textrm{ext}^\textrm{h}\left(x=x_4,t\right)&=p_\textrm{ext,R}^\textrm{h}\\
\end{aligned}
\end{equation}
\noindent and the electrode at $x=x_4$ grounded $\phi(x_4)=0$. The jumps at the solution-pore interfaces for $p^\textrm{h}$, $c_\textrm{T}$, and $\phi$ represented by Eq. (\ref{Eq:Up_model_BC}) still hold and we add the following ion flux continuity equations at $x=x_2$, $x_3$:
\begin{equation}
\begin{aligned}
j_{\text{ions,ext}}\left(t\right) &=\epsilon_\textrm{p} \cdot j_{\text{ions}}\left(t\right); \quad u_{x,\text{ext}}\left(t\right) =\epsilon_\textrm{p} \cdot u_{x}\left(t\right); \quad j_\textrm{ch,ext}\left(t\right)=\epsilon_\textrm{p}\cdot j_\textrm{ch}\left(t\right)\\
j_{\textrm{ions,ext}}\left(t\right) &= -2\cdot u_{x,\textrm{ext}}\left(t\right) c_\textrm{ext}\left(x,t\right) +2\cdot \frac{\partial c_\textrm{ext}\left(x,t\right)}{\partial x}\\ &=-u_x\left(t\right)c_\textrm{T}\left(x,t\right)+\cosh\xi\frac{\partial c_\textrm{T}\left(x,t\right)}{\partial x}\\
&-\left(\omega X_\textrm{m}\cosh\xi+c_\textrm{T}\left(x,t\right)\sinh\xi\right)\frac{\partial \phi\left(x,t\right)}{\partial x}\\
\end{aligned}
\end{equation}\par
For both Case I and II the function describing the (unbiased) current density, for a AC operation, has the form:
\begin{equation}
j_\textrm{ch}\left(t\right)= j_\textrm{ch,0} \cdot\sin\left(f\cdot t\right)
\end{equation}
where $j_\textrm{ch,0}$ is the perturbation of the current and $f$ and $f_\textrm{dim}=f\cdot D_\textrm{ref}/L_\textrm{ref}^2$ are the dimensionless and dimensional frequencies, respectively. The PDE-AE system was discretized on the $x$-direction to a differential algebraic equation system by means of the \textit{method-of-lines} with an unequal mesh grid. 
\section{Results and Discussion}
\begin{figure}
	\centering
	\includegraphics[width=0.95\textwidth, page=2]{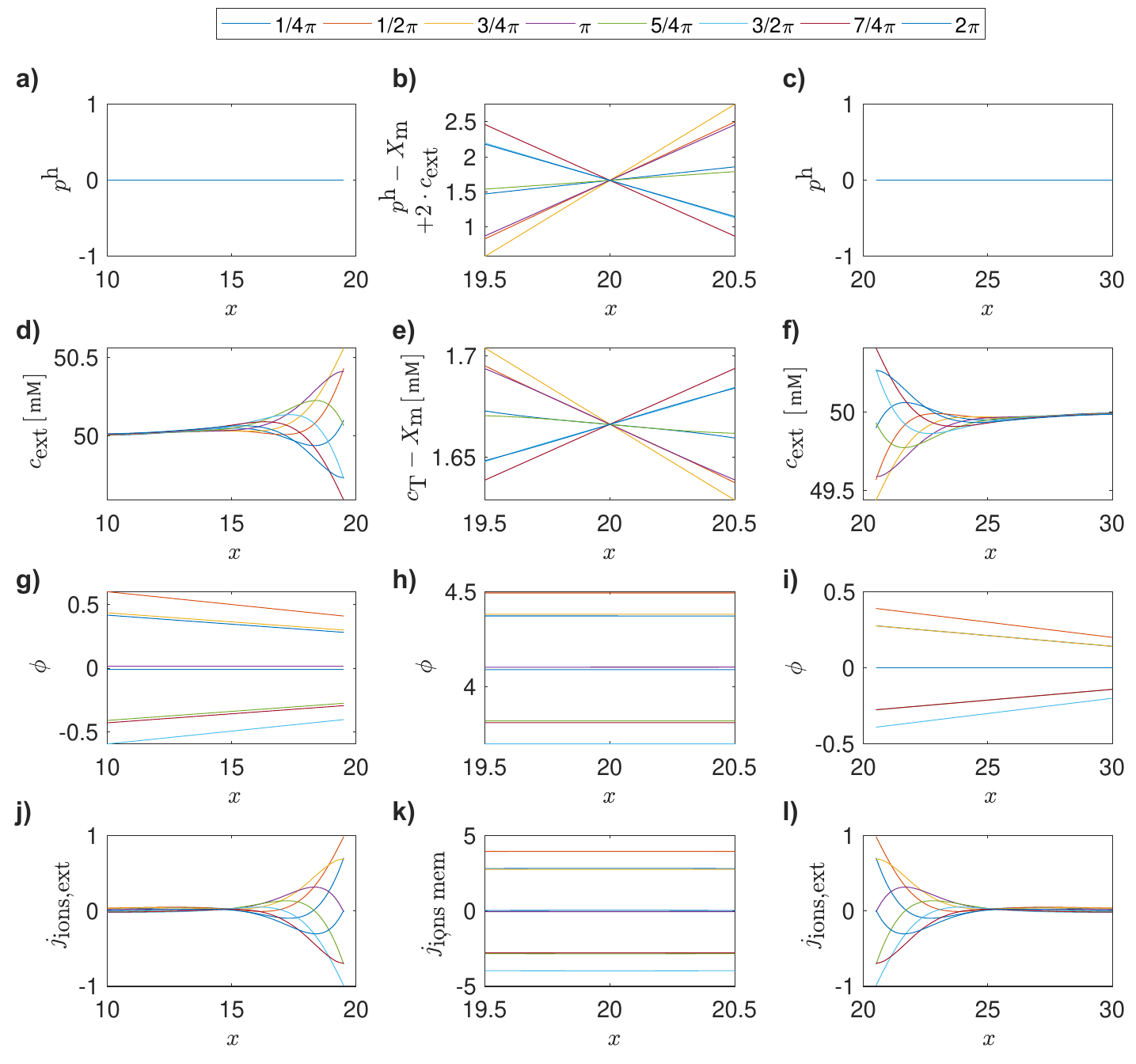}
	\caption{a), b), and c): Hydrostatic pressure; d), e), and f): Concentration; g), h), and i): Electrical potential difference; and j), k) and l): Total ion flux profiles at different  times. Frequency $f=1$ ($f_\textrm{dim}=0.2$ Hz), $X_\textrm{m}=3$ M, $\omega=+1$, and $D_{+}=D_{-}=2 \cdot 10^{-9}$ m$^2/$s.}
	\label{fig:All_profiles}
\end{figure}
In this section, we will present calculation results for one realistic scenario for the two Cases discussed above. We will use one set of parameters as discussed next. The pores have dimensional radius of $r_\textrm{p}=2$ nm, membrane charge density $X_\textrm{m} \in \left[0.3, 3\right]$ M constant in the $x$-direction, positively charged pore walls ($\omega=+1$), and porosity $\epsilon_\textrm{p}=0.25$. The solution viscosity, ion diffusion coefficient, and temperature used were, $\mu =1\cdot 10^{-3}$ Pa$\cdot$s, $D_\textrm{ref}\in \left[2\cdot 10^{-10}, 2\cdot 10^{-9}\right]$ m$^2/$s with $D_{-}/D_{+}\in\left[\nicefrac{1}{4},4\right]$, and $T=298$ K, respectively. The perturbation current density was fixed at $j_\textrm{ch,0}=1$ and we considered  $p^\textrm{h}_\textrm{ext}\left(x_1,t\right)=p^\textrm{h}_\textrm{ext}\left(x_4,t\right)$.\par
\begin{figure} 
	\centering
	\includegraphics[width=0.70\textwidth, page=3]{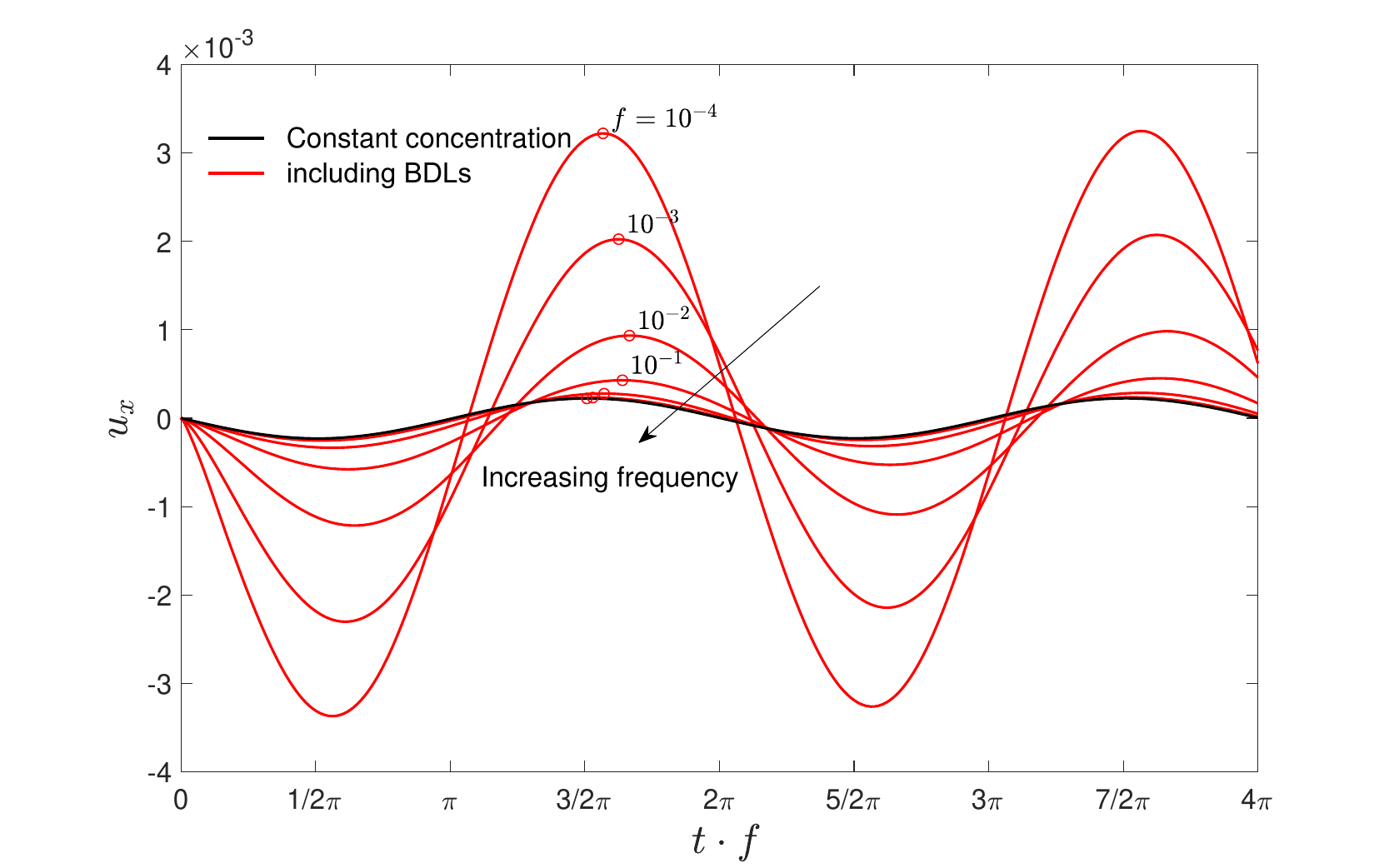}
	\caption{$u_x$ at different frequencies: Case I where the pore ends are kept at constant concentration (black line); Case II where BDLs are included (red lines). Parameters used: $c_\textrm{ext,L}=c_\textrm{ext,R}=50$ mM, $X_\textrm{m}=3$ M, $\omega=+1$, and $D_\textrm{+}=D_\textrm{-}=2\cdot 10^{-9}$ m$^2/$s.}
	\label{Fig:Ux_different_freq}
\end{figure}
Figure \ref{fig:All_profiles} shows the hydrostatic pressure, (a)-(c), concentration, (d)-(f), electrical potential difference, (g)-(i), and ion flux profiles, (j)-(l) in the three domains at different times (evenly spaced of $\nicefrac{1}{4}\; \pi$) within a single AC cycle. The frequency was fixed at $f=1$, $X_\textrm{m}=3$ M, $\omega=+1$, the diffusion coefficients $D_{+}=D_{-}=2 \cdot 10^{-9}$ m$^2/$s, and the BDLs length much larger that of the pore (giving $\left. \partial c_\textrm{ext} / \partial x\right|_{x=x_1} \sim 0 $ and $\left. \partial c_\textrm{ext} / \partial x\right|_{x=x_4} \sim 0 $). The system of equations was discretized in the flux direction ($x$) with a non-uniform grid with finer mesh at the interfaces between the pore and BDL (total 886 grid points). The initial conditions for the grid points were calculated considering the solution of the steady-state problem with $j_\textrm{ch}=0$. From Figure \ref{fig:All_profiles} it is clear that the central point of the membrane is a symmetrical point for the concentration and the hydrostatic pressure difference; thus the system considering an unbiased oscillating current and equal concentration at the external boundaries of the BLDs can be also modelled considering half of the domain.
The step changes of the total ion flux at the pore/solution interfaces come from the flux continuity equations which account for the membrane porosity.\par
\begin{figure} 
	\centering
	\includegraphics[width=0.70\textwidth, page=4]{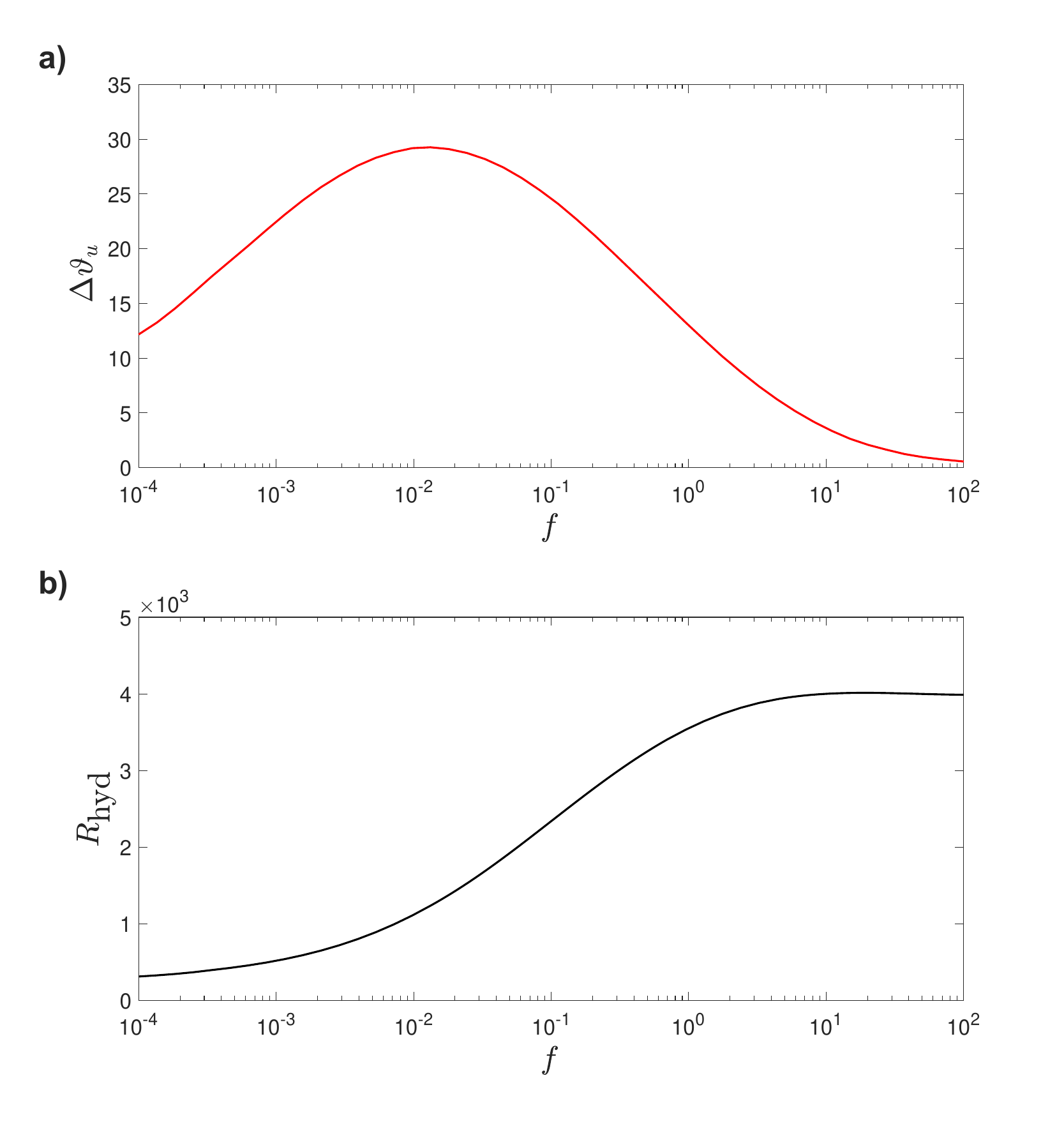}
	\caption{a): Phase shift between $u_x$ and $-j_\textrm{ch}$ at different frequencies. b): Amplitude of the EO coupling factor ($R_\textrm{hyd}= j_\textrm{ch,0}/\max \left(u_x\right)$) versus the frequency. Parameters used: $c_\textrm{ext,L}=c_\textrm{ext,R}=50$ mM, $X_\textrm{m}=3$ M, $\omega=+1$, and $D_\textrm{+}=D_\textrm{-}=2\cdot 10^{-9}$ m$^2/$s.}
	\label{Fig:Phase_shift_and_amplitude}
\end{figure}
Figure \ref{Fig:Ux_different_freq} reports the perturbation of the velocity at different applied frequencies ($f\in \left[10^{-4}, 10^{3}\right]$) for the constant pore-end concentration problem (black line, Case I) and when the BDLs are considered (red lines, Case II). In these calculations the current density flowing through the pore was the same for the two cases; the membrane charge density was fixed at $X_\textrm{m}=$3 M which is a typical value for ion exchange membranes and the solution was chosen to mimic a dilute KCL solution (50 mM) ($D_\textrm{+}=D_\textrm{-}=2\cdot 10^{-9}$ m$^2/$s). Here it is clear that in Case 1, thus when the ends of the pore have a fixed concentration, we do not see a change of the phase and neither in the amplitude of the volumetric flux. However, in Case II, where we add BDLs to the problem, we now do observe a phase shift and a modulation in the amplitude of the volumetric flux. To highlight this behaviour the phase shift ($\Delta \vartheta_u$) between $u_x$ and $-\omega j_\textrm{ch}$ and the amplitude of the EO coupling factor are reported in Figure \ref{Fig:Phase_shift_and_amplitude}a and b, respectively. The phase lag is related to the ion diffusion in the BDLs and, with the set of parameters used in these calculations, has a maximum value of $\sim\pi/6$ at $f=0.01$. When anion (cation) exchange membrane are used the volumetric flux is out of (in) phase with respect to the applied current. The EO coupling factor is defined per unit of current density as: $R_\textrm{hyd}=j_\textrm{ch,0}/\max \left(u_x\right)$ and  this value also corresponds to its dimensional form. $R_\textrm{hyd}$ increases by increasing the frequency until a plateau value is reached (for $f>10$). At even higher frequencies the AC cycle is too rapid to allow for an effective ion and solution transport through the membrane, and the efficiency of the AC pump drops steeply.\par 
Calculations with different values between $c_\textrm{ext,L}$ and $c_\textrm{ext,R}$ (not reported, $c_\textrm{ext,L}=10$ mM, $c_\textrm{ext,L}=100$ mM) showed that the current amplitude (with an average current $\textlangle{}j_\textrm{ch}\textrangle{}$=0, frequency $f=1$, and $j_\textrm{ch}\in\left[0.1,10\right]$) had no influence on the time-average value of $u_x$. Calculations were performed both considering an oscillating current density as well as an asymmetric current signal (short burst at high current in one direction, longer period at lower current in the opposite direction still with $\textlangle{}j_\textrm{ch}\textrangle{}$=0). Thus, application of either an AC or asymmetric current signal, with relatively small amplitude, between two reservoirs at different salt concentration did not modify the value of the flow of solution, from its value without current (pure osmotic flow).\par
\begin{figure} 
	\centering
	\includegraphics[width=0.70\textwidth, page=5]{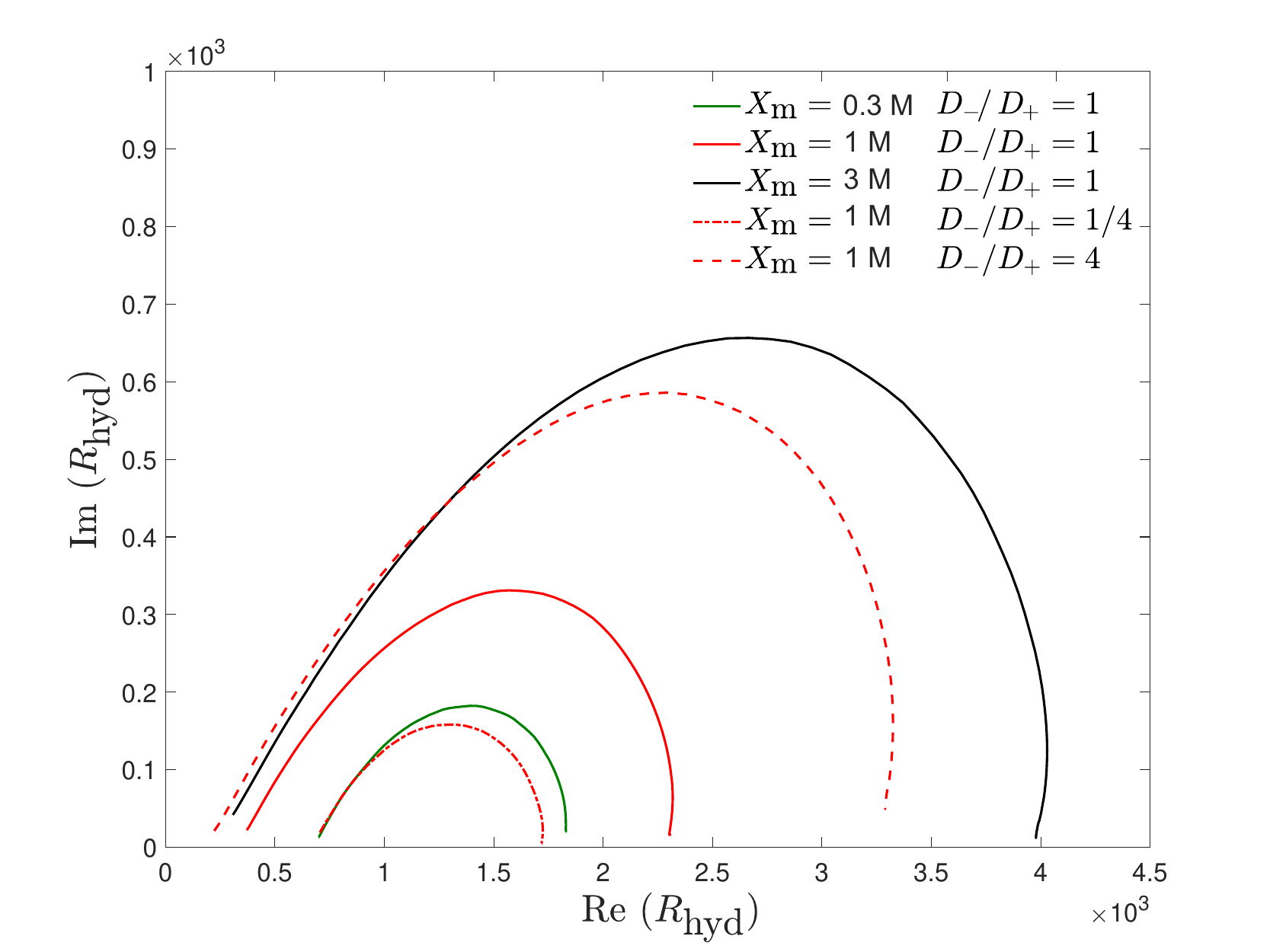}
	\caption{Nyquist plots for the EO coupling factor. Parameters used: membrane charge density of $X_\textrm{m}=$0.3, 1 and 3 M, $\omega=+1$, external concentration $c_\textrm{ext,L}=c_\textrm{ext,R}=50$ mM,  $D_\textrm{ref}=2\cdot 10^{-9}$ m$^2/$s, and $D_{-}/D_{+}\in\left[1/4,4\right]$.}
	\label{Fig:Nyquist_plot}
\end{figure}
Figure \ref{Fig:Nyquist_plot} shows the Nyquist plots for $R_\textrm{hyd}$ in pores with a membrane charge density of $X_\textrm{m}=$0.3, 1, and 3 M. By lowering the membrane charge the influence of the charged pore on the ion transport is dramatically reduced and the magnitude of the phase-lag decreases until it becomes negligible (for $X_\textrm{m}<0.1$ M, not shown in Figure \ref{Fig:Nyquist_plot}). In the same figure also a comparison among different diffusion coefficients ratios, holding constant the geometric mean $D_\textrm{ref}=2\cdot 10^{-9}$ m$^2/$s, is shown for a membrane charge density of $X_\textrm{m}=1$ M. It is evident that by increasing the diffusion coefficient of the anion (counterion with respect to the immobile charges in the pore walls) the phase lag increases. In Figure \ref{Fig:Nyquist_plot} we have reported the calculations for $\omega=+1$, representing an anion-exchange membrane, nonetheless the Nyquist plots will be identical also for cation exchange membranes (having the same charge density and pore radius) if the reciprocal of the diffusion coefficient ratio is considered.\par
Next we will evaluate how the information regarding the dynamic response of the EOF to oscillating current can be used to assess the diffusion coefficient in a true membrane. Considering a membrane with thickness of $\delta=50-100$ $\mu$m and $X_\textrm{m}=3$ M -which are the typical values for ion exchange membranes used e.g in fuel cells and redox flow batteries (i.e. Nafion 212 membranes)- the range of frequency of interest where the phase shift is observed is $f_\textrm{dim} \in \left[10^{-4},10^1\right]$ Hz when the diffusion coefficients have the characteristic values of K$^{+}$ and Cl$^{-}$. This range is of particular interest because it is inside the range of commonly used sampling frequencies for EOF experiments.\par
\begin{figure} 
	\centering
	\includegraphics[width=0.95\textwidth, page=6]{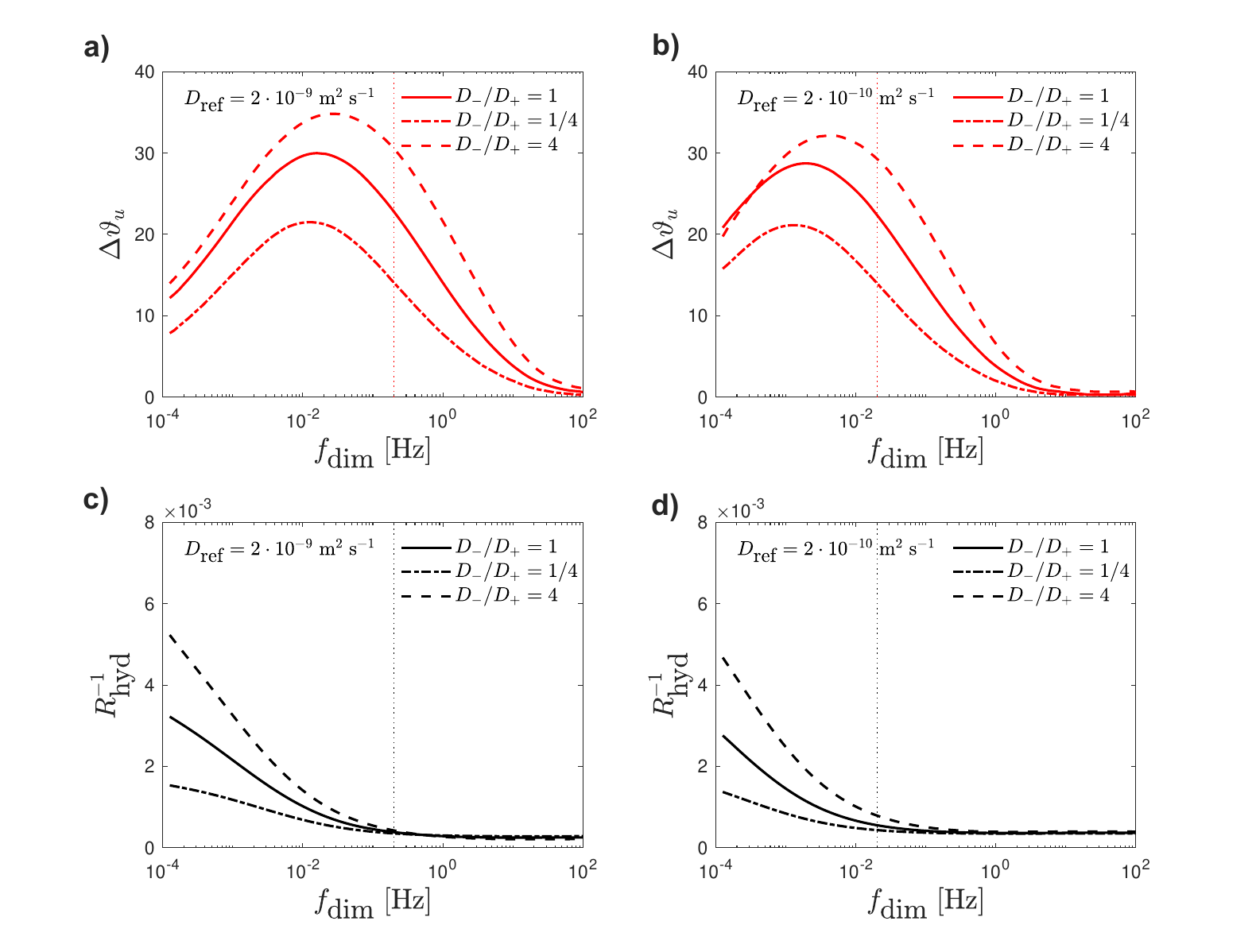}
	\caption{a) and b) Phase shift between $u_x$ and $-\omega j_\textrm{ch}$, and c) and d) amplitude of $R_\textrm{hyd}^{-1}$ for $D_\textrm{ref}=2\cdot 10^{-9}$  m$^2/$s (panels a and b) and $2\cdot 10^{-10}$ m$^2/$s (panels c and d) for $D_{-}/D_{+}=\nicefrac{1}{4},1$ and 4. The pore length is equal to $L_\textrm{ref}=100$ $\mu$m, the external concentration $c_\textrm{ext,L}=c_\textrm{ext,R}=50$ mM, pore membrane charge density $X_\textrm{m}=3$ mM, and $\omega=+1$. }
	\label{Fig:Diff_coeff_comparison}
\end{figure}
In practical terms, the analysis of the phase shift and amplitude curves gives a first order approximation of the geometrical mean and ratio of coion and counterion diffusion coefficients in the membrane, respectively. This concept is more evident when we return to the dimensional form for the frequency and we report the reciprocal of the EO coupling factor. Figure \ref{Fig:Diff_coeff_comparison} shows the phase shift and amplitude of $R_\textrm{hyd}^{-1}$ versus the dimensional frequency for ions having two different mean diffusion coefficients: $D_\textrm{ref}=2\cdot 10^{-9}$ and $2\cdot 10^{-10}$ m$^2/$s. Here it is clear that the dynamic response of the system is related to the diffusion coefficient; when the frequency equals $f_\textrm{dim}=D_\textrm{ref}/L_\textrm{ref}^2$, $R_\textrm{hyd}^{-1}$ is at the onset value. Shifting the attention to the diffusion limited regime at low frequencies we notice that the asymmetry of the absolute values of the anion and cation diffusion coefficient is reflected in the slope of $R_\textrm{hyd}^{-1}$. With reference to the bottom panels of Figure \ref{Fig:Diff_coeff_comparison} one can immediately observe that the slope of $R_\textrm{hyd}^{-1}$ versus $\log_{10} \left(f\right)$ scales with $\sqrt{D{-}/D_{+}}$. This ultimately allows one to retrieve the separate values of the diffusion coefficient for the coion and counterion.\par 
A practical device for the measurement of the diffusion coefficient from the AC-driven EOF might be similar to  the one described in detail in Ref. \cite{OSTEDGAARDMUNCK2017}. In particular a flow cell is connected to two large reservoirs containing the same electrolyte solution and one outlet-port of the low-pressure side is connected to a fine scale (sampling at frequency of the order of 10 Hz) monitoring the time evolution of the permeated volume. Additionally to conduct an AC-driven experiment, the current collectors of the flow cells need to be connected to a waveform generator or to an electrochemical workstation generally used for electrochemical impedance spectroscopy. The choice of the electrode material is dictated by the redox pair studied and, in general, can be reversible electrodes (e.g. Ag/AgCl) or carbon electrodes for most organic redox species. 
In a real experiment the apparent diffusion coefficient of the ions in the membrane is calculated from the onset of $R_\textrm{hyd}^{-1}$ increase by means of: $D_\textrm{app}=1/f_\textrm{dim,onset}\cdot\delta_\textrm{mem}^2$ where $\delta_\textrm{mem}$  represents the membrane thickness.\par
Finally we notice that we defined the diffusion coefficient as apparent (underestimated) since it is scaled with the membrane thickness and is therefore dependent on the tortuosity factor of the ion channel. However, when the Nernst-Planck framework is adopted to describe the membrane behaviour in electro-membrane processes, often the diffusion coefficient used is the apparent one (containing contributions from the porosity and tortuosity factor) see e.g. Ref. \cite{Tedesco2016}. Thus the measured apparent diffusion coefficient via the AC method might be directly adopted for membrane modelling without any further consideration regarding the ill-defined tortuosity factor of the pore network. In any event the dynamic response of the volumetric flow associated with an AC perturbation can provide insight in the diffusion coefficients of salt in highly charged membranes, and this method is envisioned to be useful for organic ions (e.g. novel redox species synthesised for all-organic flow batteries), of which kinetic properties in membranes are virtually undescribed in the literature. 

\section{Conclusions}

In this work we have derived the equations for the ion and solution transport in charged nanocapillaries driven by an oscillating current density. In particular we have considered the limiting case of the Uniform Potential model not accounting for the radial dependence of the electrical potential in the charged pore. By considering the nanoscopic pore in equilibrium with two boundary diffusion layers it is possible to describe the phase-lag of the permeated volumetric flux with respect to the applied current density. We showed the Nyquist plots for the EO coupling factor of the pore at different membrane charge densities and different values of the anion and cation diffusion coefficients. Interestingly, the phase-shift is observed in the frequency range normally used for sampling the electro-osmotic flow in separators or membranes ($f_\textrm{dim} \in \left[10^{-4},10^1\right]$ Hz). We proposed the use of the EOF-AC method to determine the ion diffusion coefficient in a configuration similar to the flow batteries cell integrated with a wave generator of an electrochemical workstation. In practical term the ion diffusion coefficient measurement is based on the analysis (\textit{i}) of the frequency at which the increase of the reciprocal of the EO coupling factor has its onset, and (\textit{ii}) its slope  versus the frequency in the low frequency range. 

\section*{Acknowledgements} 
JC gratefully acknowledges the affiliation with the Center for Integrated Materials Research (iMAT) at Aarhus University.

\newcommand{\noopsort}[1]{} \newcommand{\printfirst}[2]{#1}
\newcommand{\singleletter}[1]{#1} \newcommand{\switchargs}[2]{#2#1}

\end{document}